# Main Manuscript for

# Twitter and Facebook posts about COVID-19 are less likely to spread false and low-credibility content compared to other health topics

David A. Broniatowski[1,2,*], Daniel Kerchner[3], Fouzia Farooq[4], Xiaolei Huang[5], Amelia M. Jamison[6,†] , Mark Dredze[7], Sandra Crouse Quinn[6]

[1]Department of Engineering Management and Systems Engineering, School of Engineering and Applied Science, The George Washington University, Washington, DC 20052, USA

[2]Institute for Data, Democracy, and Politics, The George Washington University, Washington, DC 20052, USA

[3]George Washington University Libraries, The George Washington University, Washington, DC 20052, USA.

[4]Department of Epidemiology, Graduate School of Public Health, University of Pittsburgh, Pittsburgh, PA 15260, USA.

[5]Department of Computer Science, University of Memphis, Memphis, TN 38152, USA

[6]Department of Family Science, Center for Health Equity, School of Public Health, University of Maryland, College Park, MD 20742, USA

[7]Department of Computer Science, Whiting School of Engineering, Johns Hopkins University, Baltimore, MD 21218, USA

*David A. Broniatowski

**Email:** broniatowski@gwu.edu

**Author Contributions:** David A. Broniatowski: Conceptualization, Methodology, Software, Validation, Formal Analysis, Writing – Original Draft, Visualization, Supervision, Project administration, Funding Acquisition; Daniel Kerchner: Conceptualization, Software, Validation, Formal analysis, Investigation, Resources, Data Curation, Writing – Original Draft; Fouzia Farooq: Conceptualization, Software, Formal analysis, Investigation, Resources, Data Curation, Writing – Original Draft; Xiaolei Huang: Software, Investigation, Resources, Data Curation, Writing – Original Draft; Amelia M. Jamison: Investigation, Writing – Review & Editing; Mark Dredze: Software, Formal analysis, Investigation, Resources, Writing – Review & Editing, Supervision; Sandra Crouse Quinn: Writing – Review & Editing, Supervision; Project administration.

**Competing Interest Statement:** David A. Broniatowski received an honorarium from the United Nations Shot@Life Foundation – a non-governmental organization that promotes childhood vaccination. Mark Dredze holds equity in Sickweather Inc. and has received consulting fees from Bloomberg LP and Good Analytics Inc. None of these organizations had any role in the study design, data collection, ana analysis, decision to publish, or preparation of the article. The remaining authors declare no competing interests.

**This PDF file includes:**

Main Text
Figures 1 to 4

## Abstract

On February 2, 2020, the World Health Organization declared a COVID-19 social media "infodemic", with special attention to misinformation – frequently understood as false claims. To understand the infodemic's scope and scale, we analyzed over 500 million posts from Twitter and Facebook about COVID-19 and other health topics, between March 8 and May 1, 2020. Following prior work (Lazer et al., 2019), we assumed URL source credibility is a proxy for false content, but we also tested this assumption. Contrary to expectations, we found that messages about COVID-19 were *more* likely to contain links to *more credible* sources. Additionally, messages linking to government sources, and to news with intermediate credibility, were shared more often, on average, than links to non-credible sources. These results suggest that more ambiguous forms of misinformation about COVID-19 may be more likely to be disseminated through credible sources when compared to other health topics. Furthermore, the assumption that credibility is an adequate proxy for false content may overestimate the prevalence of false content online: less than 25% of posts linking to the least credible sources contained false content. Our results emphasize the importance of distinguishing between explicit falsehoods and more ambiguous forms of misinformation due to the search for meaning in an environment of scientific uncertainty.

## Significance Statement

Vast quantities of online misinformation about COVID-19 could undermine pandemic containment efforts. We analyzed hundreds of millions of social media posts to determine if the COVID-19 posts pointed to lower-credibility sources compared to other health content. We found that messages about COVID-19 contained proportionally **more** links to credible sources, with links to government websites and intermediate credibility sources more likely to go viral. Most content from the least credible sources was not explicitly false. Beyond fact-checking efforts, our findings highlight the need to address non-falsifiable content when responding to the COVID-19 infodemic.

## Main Text

### Introduction

COVID-19, an illness caused by the SARS-COV-2 virus, is a potentially fatal disease that was declared a pandemic on March 11th, 2020 by the World Health Organization (WHO). On February 2, 2020, as the pandemic was still emerging and worldwide attention became focused on the outbreak, the World Health Organization declared a COVID-19 social media "infodemic" – "an overabundance of information—some accurate and some not—that makes it hard for people to find trustworthy sources and reliable guidance when they need it" (1). Significant quantities of untrustworthy content shared online can hamper an effective public health response and create confusion and distrust among people, ultimately leading to significant loss of life. This concern led to a recent joint statement by the WHO and the United Nations (https://www.who.int/news/item/23-09-2020-managing-the-covid-19-infodemic-promoting-healthy-behaviours-and-mitigating-the-harm-from-misinformation-and-disinformation), and several other characterizing this infodemic by its propensity to propagate misinformation. Other high profile sources (2–4) have emphasized the spread of conspiracy theories, scams, and other explicit falsehoods that might lead to unhealthy behaviors and loss of life. The infodemic has thus been characterized as a "misinfodemic" (11).

To our knowledge, no systematic evidence has been proffered to support the claim that COVID-19 is more likely to spread false content than other topics. We therefore seek to characterize the infodemic's scale and scope. Furthermore, if the infodemic is indeed characterized by false content, one might expect a higher proportion of this content to come primarily from sources that "lack the news media's editorial norms and processes for ensuring the accuracy and credibility of information" (12). This should be especially true when comparing COVID-19 social media samples to equivalent samples pertaining to other topics. Prior estimates of the prevalence of credible content about COVID-19 (13–16) both vary widely and have not been compared to equivalent samples from other time periods, or other topics. Cross-platform comparisons are few in number (16). Thus, our first question is:

(1) Are posts about COVID-19 more likely to contain links to low-credibility sources when compared to other health topics?

To answer this question, we analyzed hundreds of millions of posts shared on the two largest social media platforms in the USA – Facebook and Twitter – and examined the volume and source credibility of URLs shared in these posts.

Beyond the sheer volume of links shared, one might define an "infodemic" by the likelihood that a particular type of post might go viral. Thus, our second question is:

(2) When it comes to COVID-19, were links to low-credibility news sources more likely to go viral?

Like prior work (12, 14, 16–19), our analysis draws upon a widespread simplifying assumption: that "the attribution of 'fakeness' is … not at the level of the story but at that of the publisher." (17). This assumption is attractive because it is scalable, allowing researchers to analyze vast quantities of posts by characterizing their sources. If this assumption is true, we would expect most content from low-credibility sources to contain false claims. Thus, our third question is:

(3) When it comes to COVID-19, did most links to low-credibility news sources contain false claims?

There is significant variability within the broad category of misinformation, ranging from blatantly false conspiracy theories to claims that were once considered plausible but have been disproven. To better understand how source credibility varies with topic, we consider two COVID-19 misinformation topics: 1) the conspiracy theory that 5G wireless technologies cause COVID-19 – a conspiracy theory selected because it has no basis in scientific fact (https://news.un.org/en/story/2020/04/1062362); 2) the claim that hydroxychloroquine is an effective COVID-19 treatment – selected because it reflected a case of genuine scientific uncertainty during the time period of our study (March – May, 2020), when leaders within the scientific community expressed caution, but no clinical trials had yet been conducted (20, 21). (Clinical trials have so far failed to demonstrate hydroxychloroquine's efficacy; https://www.nih.gov/news-events/news-releases/nih-halts-clinical-trial-hydroxychloroquine. Nevertheless, WHO "mythbusters" currently states that "more decisive research is needed to assess [hydroxychloroquine's] value in patients with mild disease or as pre- or post-exposure prophylaxis in patients exposed to COVID-19."; https://www.afro.who.int/health-topics/coronavirus-covid-19/mythbusters).(22, 23). Thus, our fourth question is:

(4) When it comes to specific COVID-19 narratives pertaining to 5G wireless and hydroxychloroquine, which types of sources were more likely to spread which types of misinformative content?

Beyond source credibility, several reports have alleged that state-sponsored actors used the infodemic to promote narratives advancing their geopolitical interests (6–8). If so, we would expect to see a relative increase in the number of links to state-sponsored propaganda outlets. Thus, our fifth question is:

(5) Was the proportion of links to state-sponsored sources larger for COVID-19 than for other health topics?

## Results

**Dataset Statistics.** We identified 599,745,125 posts on Twitter, Facebook Pages, and Facebook Groups containing keywords pertaining to COVID-19, vaccine-preventable illnesses, and other health conditions. These posts contained 169,201,404 URLs including 762,669 unique top-level domains (TLDs; see Table S1). 16,614 (2%) of these TLDs were assigned a credibility rating using NewsGuard or MediaBiasFactCheck scores, these TLDs accounted for 48% of all URLs shared after removing in-platform links (i.e., retweets and shares; Figure 1). An additional 80,539 (11%) TLDs were unrated but assigned a category by WebShrinker. More than 90% of URLs pointed to one of these categorized or rated TLDs.

**Posts about COVID-19 are more likely to contain links to high-credibility sources when compared to other health topics.** Across all datasets, quantities of "More Credible" domains exceeded "Less Credible" and "Not Credible" domains combined. Furthermore, compared to other health topics, COVID-19 datasets contained the highest proportions of rated URLs and, among these, the highest proportions of "More Credible" URLs (all $p<0.001$; Table S2, Figure 2).

**The least credible posts are not the most viral.** In all COVID-19 datasets, posts containing URLs linking to "not credible news" were never the most viral category. Compared to these posts, Government Website URLs were more viral on Twitter, $B=0.51$, $p<0.001$, and Facebook Pages, $B=0.07$, $p<0.001$. Additionally, middle range "less credible" news sources were shared significantly more than "not credible" news sources on both Twitter, $B=0.14$, $p<0.001$, and in Facebook Groups, $B=0.45$, $p<0.001$ (Table S3).

**A minority of posts to the least credible sources contained false content.** Most claims from "not credible news" sources were not false, although these claims were more likely to contain explicit falsehoods when compared to other categories (Figure S3). In a hand-annotated stratified random sample of 3000 posts, only 303 posts (10%) contained a reference to a false claim. However, among those with "not credible" URLs, 145 (24%) contained false claims, a significantly higher rate, $\chi^2(1)=159.72, p<0.001$. Results replicated when restricting analysis to 791 posts labeled as "Information/News", $\chi^2(1)=12.78, p<0.001$.

**More credible sources.** In general, sources rated as more credible shared news and government announcements. Content was rarely political, although users posting links to these sources sometimes editorialized, often with liberal bias. Here, misinformation reported on, and potentially amplified, questionable content, such as explaining the 5G conspiracy theory or reporting on claims that bleach cures COVID. Similarly, "More Credible" URLs make up the plurality of links with posts containing keywords about hydroxychloroquine, although rates of "less credible" and "not credible" URLs were also elevated (Figure 3; $p<0.001$ in all cases, see Table S3). Some content also expressed uncertainty around COVID-19 science, pointing out limitations of data and models, and acknowledging major questions could not yet be answered.

**Less credible sources.** This intermediate category contained a variety of content (see Figure S4). Non-US politics were common, especially from Indian, Chinese, and European sources. Misinformation in this category included some political conspiracy theories, but also more subtle falsehoods including suggesting COVID is less severe than flu, promoting hydroxychloroquine as a cure, or claiming that "lockdowns" are an overreaction. This category also includes posts that attempted to debunk, but may have also amplified, false content.

**Not credible sources.** These sources contained the largest proportion of false content. Common themes included: blaming China for the virus, questioning its origins, rejecting vaccines, and framing COVID as undermining President Trump. Here, content emphasizing scientific uncertainty suggested that response measures were unjustified or that science was distorted for political ends. Propaganda narratives often extolled Russian and Chinese COVID responses, but also reported on other news events. In general, logistic regression results showed that these sources, $B=1.12$, $p<0.001$, sources containing conservative politics in general, $B=1.37$, $p<0.001$, and sources emphasizing scientific uncertainty, $B=2.82$, $p<0.001$, were more likely to contain false content.

**Unrated and in platform sources.** Figure 1 shows that the majority of URLs in any dataset were either unrated or links to other posts within the same social media platform. Inductive analysis shows that **unrated** and **in platform** content was similar, although unrated content included more links to businesses or online marketplaces (Figure S4). In both categories, lifestyle updates included event cancellations, travel restrictions, or updated business information. General discussion content included first person COVID-19 narratives, and personal opinions about news and politics. Although the overall prevalence of misinformation in these categories is low (Figure 2), both categories included opportunistic posts using the COVID-19 narrative to sell products – especially putative COVID treatments or cures – or promote events. Finally, some posts shared conspiracist content linking to more detailed information. For example, Figure 4 shows that the plurality of URLs in posts discussing 5G were unrated ($p<0.001$ in all cases; see Table S3). Many of these URLs linked to other social media platforms, "arts & entertainment" sites (especially podcasts) and "unknown" domains. Notably, the proportion of "not credible" URLs discussing 5G increased compared to other COVID-19 data, whereas the proportion of "more credible" and "less credible" URLs decreased.

**Tweets about COVID-19 were more likely to link to state-sponsored sources.** On Twitter, 12 (60%) state-sponsored domains showed elevated levels of COVID content compared to health and vaccine content, of which 7 (58%) were scored as "not credible" or "less credible" (see Table S5). These domains also had larger shares of the COVID dataset relative to both the health and vaccine datasets, when compared to the top non-state-sponsored domains ($F(1,84)=26.21$, $p<0.001$ for health, and $F(1,84)=16.18$, $p<0.001$, for vaccines). A significant two-way interaction showed that non-credible state-sponsored sources (in our dataset, from Russia, Iran, and China) also had larger shares of the COVID dataset than the vaccine dataset, $F(2,83)=4.27$, $p=0.02$, even though "not credible" URLs made up a smaller share of the COVID-19 dataset overall, $F(2,83)=4.00$, $p=0.03$ and $F(2,83)=6.67$, $p=0.003$ compared to health and vaccines, respectively. These results did not replicate across other platforms (Figure 4; see Table S6).

## Discussion

We are the first to compare the COVID-19 infodemic to other health topics across platforms. Like previous studies (14, 15, 18), we find that there is indeed an overwhelming amount of content pertaining to COVID-19 online. Although the COVID-19 "infodemic" has been widely characterized as a vector for false content from low credibility sources, we found that links to credible sources were <u>*more*</u> widespread for COVID-19 than for other health topics. Furthermore, links to government sources were more viral than links to non-credible sources on both Twitter and Facebook. Finally, the majority of posts linking to non-credible sources were not false (although posts in this category were more likely to contain falsehoods than posts in other categories). This implies that attempts to characterize the information environment using low-credibility sources as a proxy for false claims may <u>*over*</u>estimate the prevalence of these claims. When combined with the finding that sources rated as "not credible" were less likely to appear in the COVID-19 datasets, our results suggest that attempts to combat misinformation using "mythbusters" (https://www.who.int/emergencies/diseases/novel-coronavirus-2019/advice-for-public/myth-busters) or otherwise debunking false claims possibly may address only a small sliver of the infodemic.

Beyond explicitly false claims, our results suggest a large role for ambiguity, uncertainty, and attention in the infodemic. For example, credible news sources with large audiences may drive attention to misinformative content, as demonstrated by the fact that the plurality of URLs in posts discussing hydroxychloroquine – a putative cure for COVID-19 that WHO has labeled a "busted myth" (https://www.afro.who.int/health-topics/coronavirus-covid-19/mythbusters#block-hydroxychloroquine) – pointed to *more credible* news sources. This is, in part, due to scientific uncertainty around this topic. Preliminary results seemed to support hydroxychloroquine's efficacy (20, 21), whereas later, more rigorous, trials did not (23, 28). Thus, according to some definitions (29), stories about hydroxychloroquine's efficacy were not, strictly speaking, misinformation at the time our data were gathered. Nevertheless, several sources promoted anecdotal evidence overstating its efficacy (likely associated with elevated rates of lower-credibility URLs), whereas others, including major political figures, promoted hydroxychloroquine as a "miracle cure" drawing extensive media attention and attempts to debunk this claim. In reporting on this topic, mainstream news sources may have amplified a claim that was ultimately proven false (30), but which social media platforms would have had difficulty defending as misinformative at the time. This illustrates how discussion of misinformative content may spread through credible channels. these sources may intend to debunk, fact-check, or otherwise validate this content, inadvertently driving attention to it (30).

The role of ambiguity is further underscored by our finding that URLs with *intermediate* credibility scores were most viral on Twitter and in Facebook Groups. Content from this category was generally factual, yet presented in an opinionated manner, making it difficult for users to distinguish between fact, opinion, misinformation, and satire. Additionally, dubious claims may frequently arise from outside the news ecosystem. Several URLs – including the plurality of those discussing a conspiracy about 5G wireless technologies and COVID-19 – pointed to user-generated content (social media platforms and podcasts), which cannot be assigned a credibility score. Neither credibility ratings nor fact-checkers can be deployed at the scale required to assess all such content. Furthermore, there is no legitimate expectation that user-generated content is subject to journalistic fact-checking standards or academic peer review.

State-sponsored propaganda also appears to trade on this ambiguity: COVID-19 tweets were more likely to contain links to lower-credibility state-sponsored media, which prior authors have associated with harmful propaganda (5, 31). Clearly, COVID-19 was of more importance to some governments than were other health topics, consistent with previous work showing how states weaponize true and false information to promote geopolitical aims . These sources often shared content that was not, strictly speaking, factually inaccurate, although it may have promoted a narrative with geopolitical overtones (e.g., that the US is a declining power, or by critiquing flaws in Western societies). Notably, these results seem localized to Twitter. One possible explanation for this difference may be the extent to which Twitter affords artificial amplification compared to Facebook (30).

Although credibility ratings are important tools for characterizing the information environment, several sources remain to be categorized, highlighting the fact that misinformation often travels through channels outside of the news ecosystem. The swiftly-evolving online environment with significant user-generated content therefore calls into question the coverage of automated techniques based solely on source (31). Even among news sources, several non-Western, and especially African, websites were unrated, reflecting an opportunity for rating systems to expand to new media markets.

**Limitations**

Measures of URL prevalence are an indicator of information availability rather than information consumption. Thus, we cannot claim that data from social media are representative of the general public. Indeed, news sources may be more likely to be shared by automated accounts ("bots") than other types of tweets, especially when those bots serve as news aggregators (13). Furthermore, our study does not account for the fact that links to low credibility sources may sometimes be shared insincerely or sarcastically. However, to the extent that this may occur, these posts may have inadvertently amplified these sources, contributing to the overall confusion of the infodemic. Our inclusion criteria for social media data are based on keywords associated with COVID-19, vaccine-preventable illnesses, and other health conditions. This collection procedure might introduce some noise in our dataset, for example if online actors exploited the virality of the COVID-19 hashtags/keywords to promote their content. However, these opportunistic posts are a key feature of the "infodemic". Indeed, our inclusion criteria are consistent with the very definition of the infodemic. For example, a WHO/PAHO fact sheet from May 1, 2020 (https://iris.paho.org/bitstream/handle/10665.2/52052/Factsheet-infodemic_eng.pdf?sequence=14&isAllowed=y), defines the "infodemic" using keyword search terms that are similar to ours. Other studies of the "infodemic" have taken the same approach (15, 18). This approach will necessarily contain content that is not, strictly speaking, health information, it does meet WHO's definition of "a large increase in the volume of information associated with a specific topic and whose growth can occur exponentially in a short period of time due to a specific incident, such as the current pandemic. In this situation, misinformation and rumors appear on the scene, along with manipulation of information with doubtful intent." (https://iris.paho.org/handle/10665.2/52052). Nevertheless, to address this point, we included an "opportunistic" category in our annotation scheme. Our results show that this opportunistic content made up only a small percentage of overall content, with opportunistic links primarily found within the "in platform" and "unrated" categories, as expected. More broadly, this finding speaks to the difficulty of automatically rating all URLs.

**Conclusions**

Explicitly false claims about health appear a small, yet significant, feature of online health conversations. However, our results reflect a complex information environment that can't be characterized by simple "false" vs. "true" dichotomy of online content (32, 33). Although prior work has shown that the proportion of information from low credibility sources is quite small, our results demonstrate that these proportions are even smaller for COVID-19 tweets. Thus, attempts to combat misinformation that focus on low credibility sources may be missing the larger body of misinformation that is factually mixed, or even true, yet out of context, or that has changed significantly over time. Under pandemic conditions, source credibility may be a false signal of misinformative content. Beyond these considerations, we find that most content is unrated. Even if we were able to rate all sources according to meaningful credibility metrics, news aggregation sites or social media platforms that disseminate user-generated content do not employ fact-checkers at the scale required to provide credibility assessments for all of their content. Furthermore, there is no legitimate expectation that user-generated content is subject to the same fact-checking standards as journalistic, government, or academic sources. Nevertheless, this content is overwhelming in scope and varied in style, making it difficult for users to distinguish between fact, opinion, misinformation, and satire. This further underscores the importance of distinguishing between disinformation -- content that is generated for malicious purposes by non-credible actors -- and misinformation that may come about due to the search for meaning in a legitimately uncertain environment.

During an emerging pandemic, scientific findings lag public demand for answers. As perceptions of what is true change, there is an urgent need to manage expectations that public health recommendations may change without eroding trust. In contrast, fact-checkers and "mythbusters" tend to address settled knowledge and clear falsehoods. Although these tools are an important component of the scientific community's response to the pandemic, they appear not to address some of the most viral content that trades on uncertainty and ambiguity.

Although some may expect social media platforms to filter out or downrank misinformation our work shows that current methods based solely on credibility ratings may remove factual content and miss false content. Furthermore, platforms may not be equipped to respond to content that trades on uncertainty, since they evade rationales for removing explicit falsehoods. Thus, addressing the spread of misinformation during an infodemic cannot be the sole provenance of social media platforms; rather, it requires a coordinated effort between platforms, responsible news providers, public health officials, and government spokespeople. Our research points to the urgent need for interdisciplinary collaboration between computational scientists, and health, social, and communication scholars, to solve the problem of providing gists (34) – informed, contextualized human insights – to the public at Internet scale during a pandemic.

## Materials and Methods

**Data Collection.** We collected English language tweets from Twitter matching keywords pertaining to COVID-19 (35), generalized health topics (36), and vaccine-preventable illnesses between March 7, 2020 and May 1, 2020 (37) (see Supplementary Material for keywords). Health and vaccine tweets were collected for the same dates in 2019, yielding 5 datasets overall. We next collected comparable data from Facebook Pages, and Facebook Groups for the same dates using CrowdTangle (38), yielding 10 more datasets. COVID-19 Facebook posts were collected on June 2, 2020, and Facebook posts for vaccine and health keywords were collected on July 13-14, 2020.

**Credibility Categorization.** We extracted all Uniform Resource Locators (URLs) in each post. We identified each URL's top-level domain (TLD; (39)), unshortening links (e.g., "bit.ly/x11234b") if necessary (40). We enumerated the frequency of each TLD in each dataset. Next, we assigned each TLD a credibility rating between 0-100 using data collected July 5, 2020 from the NewsGuard database (https://www.NewsGuardTech.com/) and the MediaBiasFactCheck website (https://mediabiasfactcheck.com/) averaging these scores if a TLD was rated by both. TLDs with credibility scores of 66.67 or higher were rated as "More Credible", whereas those scored 33.33 or lower were rated as "Not Credible". Government (e.g., .gov) and academic (e.g., .edu) websites were assigned to the "More Credible" category and sites rated by MediaBiasFactCheck as "Questionable" were rated as "Not Credible" if they did not have a numerical rating. TLDs scoring between 33.33 and 66.67 were rated as "less credible". The remaining TLDs were marked as "unrated". Among these unrated TLDs, we labeled digital platforms (e.g., YouTube) and satire websites as such, and used the Webshrinker Category API (https://docs.webshrinker.com/) to assign the remaining domains into discrete content categories until at least 90% of all URLs were assigned a label (not counting in-platform URLs: e.g., links to facebook.com from Facebook or twitter.com from Twitter). The remaining domains were labeled as "Not categorized".

**Virality Analysis.** We conducted negative binomial regressions for each COVID dataset to predict the number of shares or retweets for each post (Facebook and Twitter share counts were current as of June 2, 2020, and May 31, 2020, respectively). Retweets were removed, since retweets report the retweet count of the original tweet. We used subcategories of TLD within the same credibility category (e.g., government vs. more credible news) as predictors since our sample size enabled us to differentiate between them. Facebook posts could have at most one URL each, but Twitter posts could have multiple URLs (although the modal post only had one URL, see Supplementary Material). We thus assigned each tweet to the credibility category of its least credible URL (see Supplementary Material).

**Qualitative Analysis.** To determine the content of each credibility category, we developed a codebook (see Supplementary Material) to assess the presence of false claims, scientific uncertainty, partisan bias, and one of five inductively-inferred content categories (politics, opportunistic, lifestyle/social, information/news, and general discussion). We next randomly selected 200 posts from each COVID-19 dataset for each credibility category (More, Less, and Not Credible, and Unrated). We also annotated 200 posts with in-platform links as a baseline for each platform, yielding a total of 3000 posts. Three authors (DK, FF, and AMJ) manually labeled batches of 100 posts each until annotators achieved high interrater reliability (Krippendorff's a>0.8, see Supplementary Materials), which we obtained on the second round (a=0.811). Disagreements were resolved by majority and ties adjudicated by a fourth author (DAB). The remaining 2400 posts were then split equally between all three annotators. We also generated qualitative descriptions for each credibility category. We examined the prevalence of false claims and scientific uncertainty in each credibility category using distribution-free chi-square tests and logistic regressions (see Table S4).

**Analysis of Hydroxychloroquine and 5G Posts.** For each topic, we calculated the proportion of all URLs in each credibility category and compared these proportions to all remaining URLs in the COVID-19 datasets using distribution-free chi-square tests.

**Analysis of State-Sponsored Propaganda.** We examined whether state-sponsored propaganda was more likely to be shared in the context of COVID-19 when compared to other health topics. We used the NewsGuard database to identified the TLDs of 37 state-sponsored news outlets, representing 13 different nations. We also generated a comparator dataset of 51 non-state-sponsored news TLDs that were among the top ten most frequently shared TLDs in at least one of the three credibility rating categories for the three COVID datasets. (3 government public health websites also appeared in the top ten "more credible" TLDs, but were excluded because they were not news sites). For each TLD, we computed a log odds ratio capturing how much more likely a TLD was to be represented in a platform's COVID-19 dataset compared to the corresponding health and vaccine datasets (see Supplementary Material). We next used distribution-free chi-square tests to determine if URLs from state-sponsored sources made up a larger share of the COVID-19 datasets compared to other datasets. To rule out the possibility that these differences were due to increased reporting on COVID-19 overall, we conducted two-way ANOVAs to determine how these log-odds ratios varied with state-sponsorship (compared to top non-state-sponsored news) and credibility category.

## Acknowledgments

This work supported in part by grant number R01GM114771 to D.A. Broniatowski and S.C. Quinn, and by the James S. and John L. Knight Foundation to the GW Institute for Data, Democracy, and Politics.

## Competing interests

David A. Broniatowski received an honorarium from the United Nations Shot@Life Foundation – a non-governmental organization that promotes childhood vaccination. Mark Dredze holds equity in Sickweather Inc. and has received consulting fees from Bloomberg LP and Good Analytics Inc. None of these organizations had any role in the study design, data collection, ana analysis, decision to publish, or preparation of the article. The remaining authors declare no competing interests.

## Ethics

The data used in this article are from publicly available online sources, the uses of which are deemed exempt by the George Washington University institutional review board (180804).

## Data Availability

All processed data used to generate the results and figures reported in this paper are available in the interactive plots included in the supplementary materials. These data are derived from five primary sources: 1) Twitter; 2) CrowdTangle; 3) MediaBiasFactCheck; 4) NewsGuard; and 5) Webshrinker. Each of these data sources are governed by their own Terms of Service. Instructions for accessing data from each source is provided in the "Data Availability" section of Supplementary Materials.

## Figures and Tables

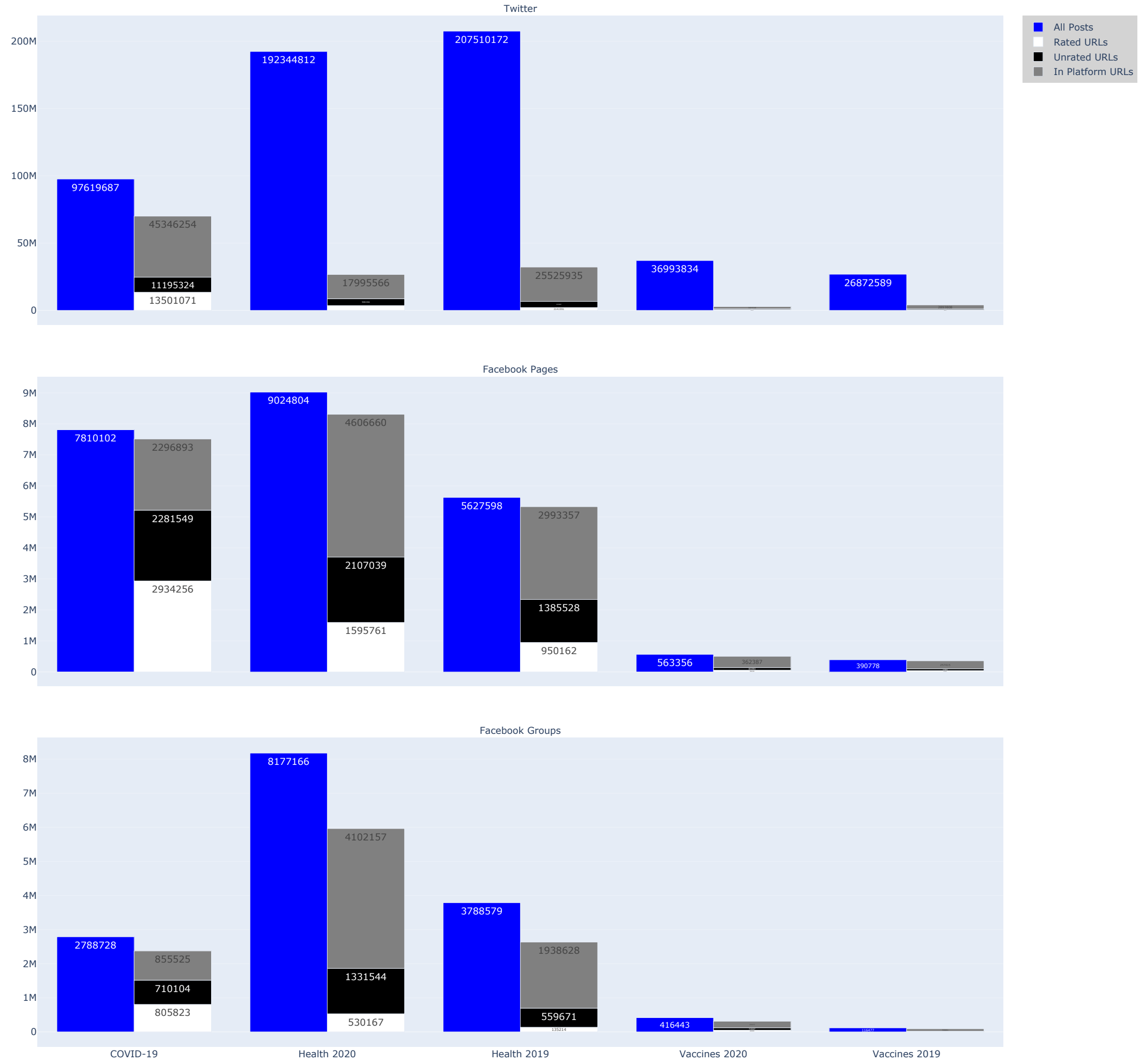

**Figure 1.** Frequency of posts (left) and URLs shared (right) in each dataset. See Figure 2 for a more detailed breakdown of out-of-platform URLs. An interactive version of this figure is available in the attached file: Figure_1_extended.html

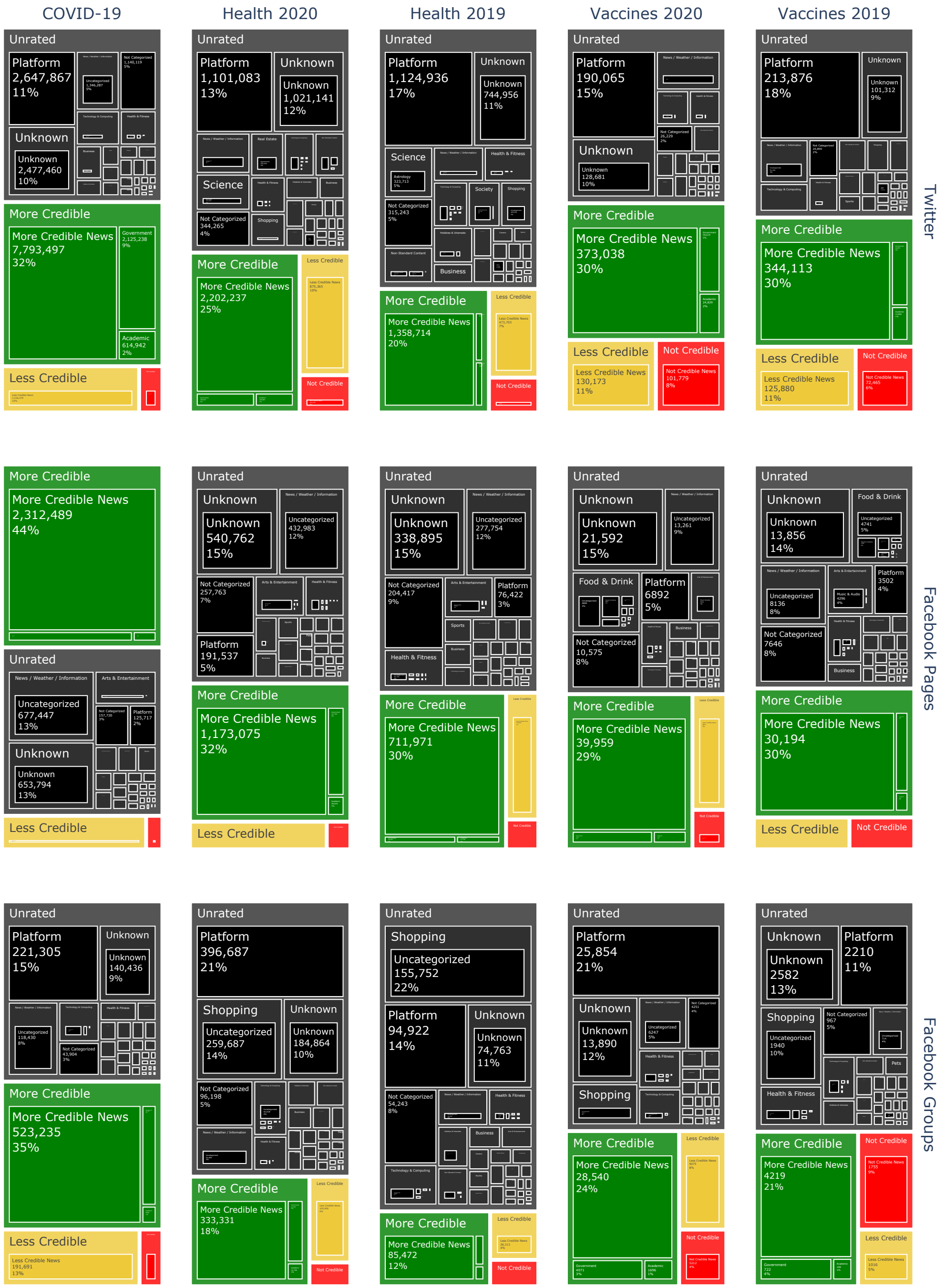

**Figure 2.** Treemaps reflecting the number of URLs for each credibility rating and category in each dataset. An interactive version of this figure is available in the attached file: Figure_2_extended.html

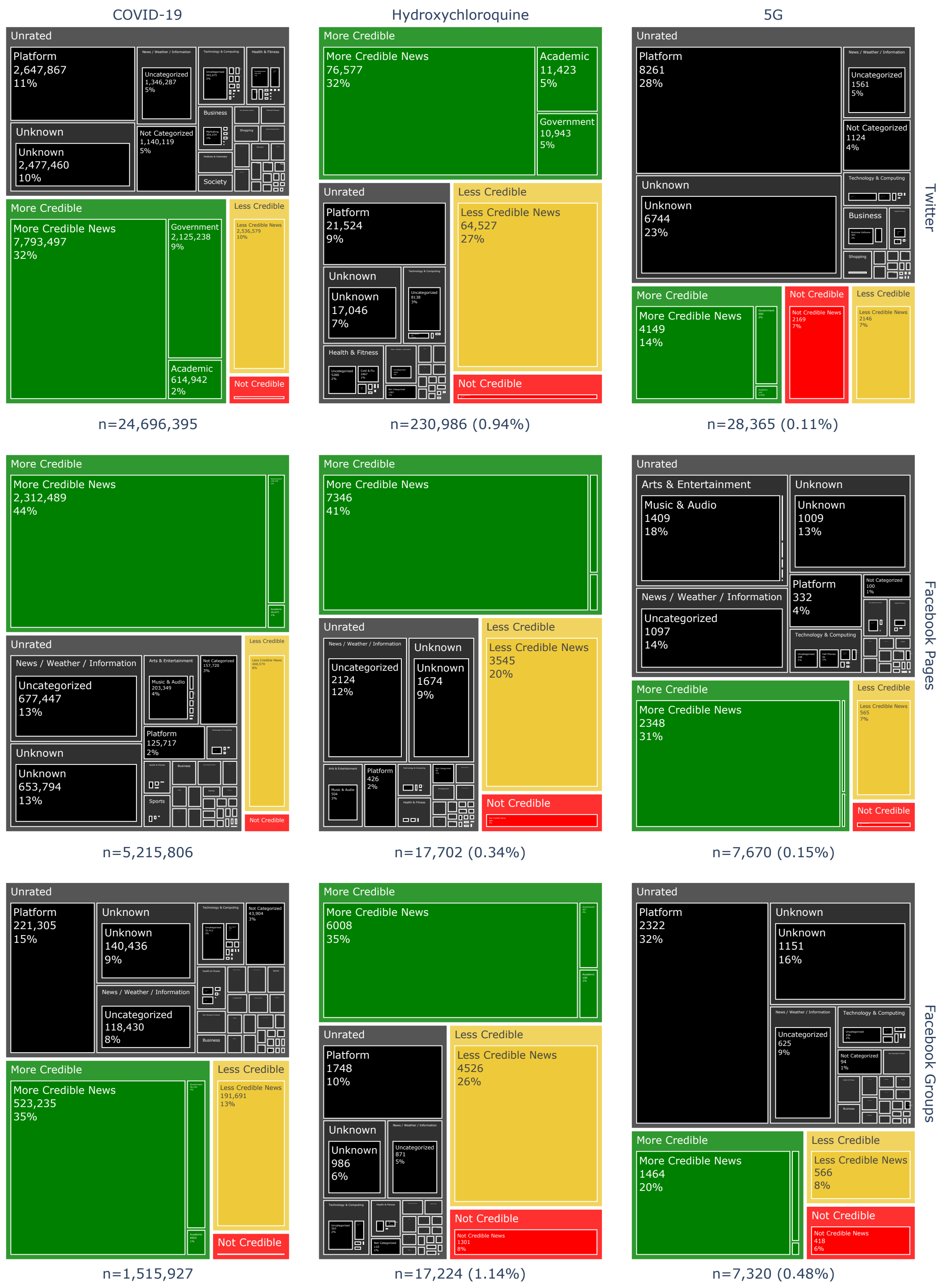

**Figure 3.** Treemaps showing proportions of hydroxychloroquine and 5G posts. All COVID-19 post proportions are also shown for comparison. An interactive version of this figure is available in the attached file: Figure_3_extended.html

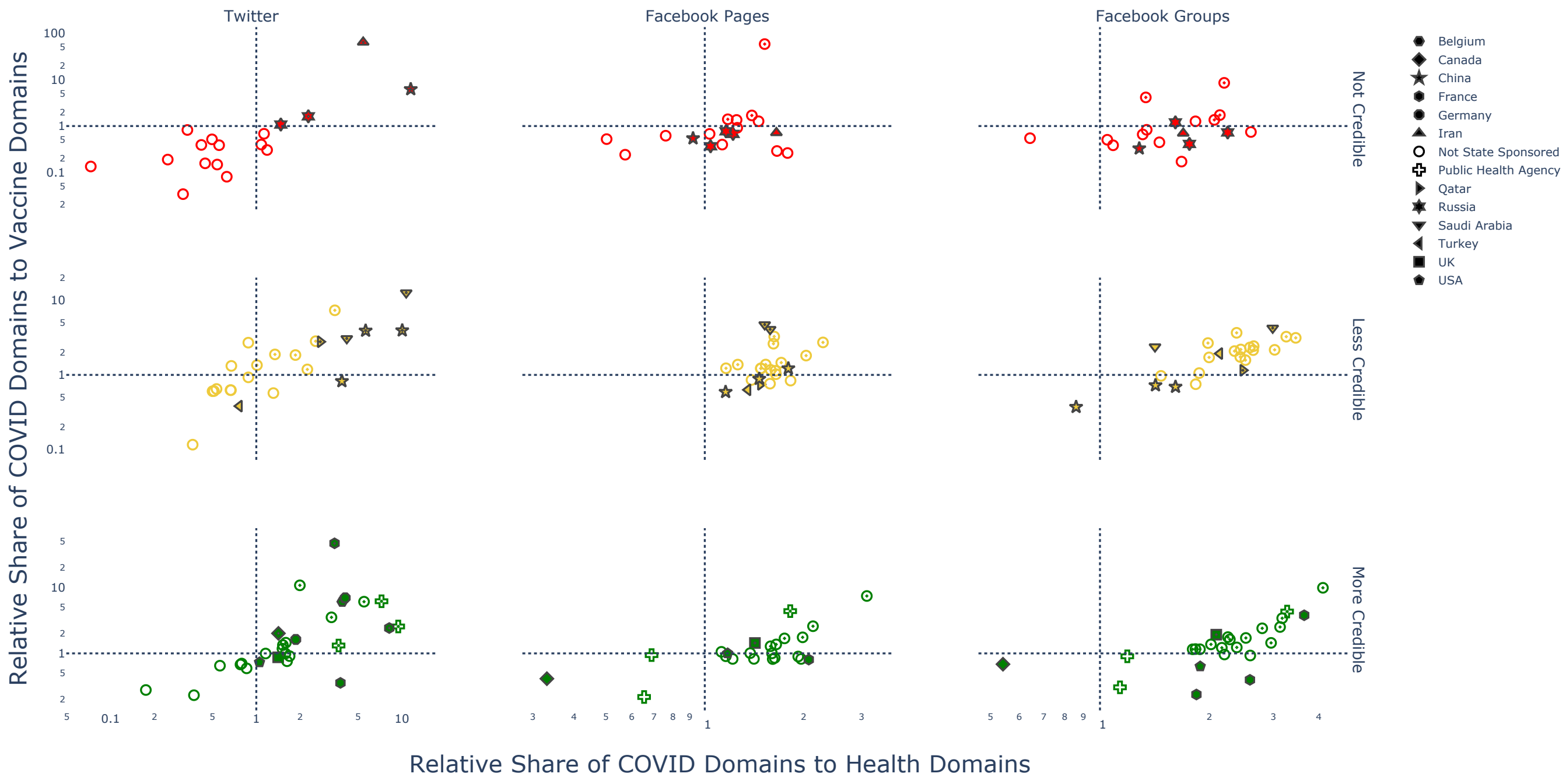

**Figure 4.** Each point represents one top-level domain. A dot inside the point indicates that the corresponding domain's share of tweets was significantly larger for COVID-19 than for both the health and vaccine datasets. Public health agencies are included for visual comparison but were not included in statistical analyses. An interactive version of this figure is available in the attached file: Figure_S4_extended.html